\documentclass[10pt,aps,prl,reprint,showkeys,%
showpacs,preprintnumbers]{revtex4-1}

\usepackage{amsmath,mathrsfs, amssymb}

\begin{document}
\bibliographystyle{apsrev4-1}

\title{Geometrical CP Violation}

\author{Ivo \surname{de Medeiros Varzielas} $^{1,2}$}
\email{ivo.de@udo.edu}

\author{David \surname{Emmanuel-Costa} $^{1}$}
\email{david.costa@ist.utl.pt}

\affiliation{$^{1}$ Departamento de F\'{\i}sica and
Centro de F\'{\i}sica Te\'orica de Part\'{\i}culas (CFTP)\\
Instituto Superior T\'ecnico, Av. Rovisco Pais, 1049-001 Lisboa, Portugal}
\affiliation{$^{2}$ Fakult\"{a}t f\"{u}r Physik, Technische Universit\"{a}t
Dortmund D-44221 Dortmund, Germany}

%11.30.Er 	Charge conjugation, parity, time reversal, and other discrete
%symmetries
%11.30.Hv 	Flavor symmetries
%12.15.Ff 	Quark and lepton masses and mixing (see also 14.60.Pq Neutrino
%mass and mixing)
%12.15.Hh 	Determination of Cabibbo-Kobayashi & Maskawa (CKM) matrix
%elements
%12.60.-i 	Models beyond the standard model (for unified field theories,
%see 12.10.-g)
%12.60.Fr 	Extensions of electroweak Higgs sector

\preprint{CFTP/10-007, DO-TH 11/19}

\date{\today}

\begin{abstract}
Spontaneous CP-violating phases that do not depend on the parameters
of the Higgs sector - the so-called calculable phases - are investigated. The
simplest realization is
in models with 3 Higgs doublets, in which the scalar potential is invariant
under non-Abelian symmetries. The non-Abelian discrete group
$\mathsf{\Delta(54)}$ is shown to lead to the known structure of
calculable phases obtained with $\mathsf{\Delta(27)}$. We investigate
the possibily of accommodating the observed
fermion masses and mixings.
\end{abstract}

\maketitle

Since the discovery of CP violation in 1964, its origin remains a fundamental
open question in particle physics. In the context of the Standard Model (SM), CP
symmetry is explicitly broken at the Lagrangian level through complex Yukawa
couplings which lead to CP violation in charged weak interactions via the
Cabibbo-Kobayashi-Maskawa (CKM) matrix. Among many mechanisms that generate CP
asymmetry beyond the SM, the possibility that CP is spontaneously broken
together with the gauge symmetry group is a very attractive
scenario~\cite{Lee:1973iz,Branco:1979pv}. One remarkable phenomenological
implication of Spontaneous CP violation (SCPV) is that it provides an appealing
solution to the strong CP problem~\cite{Mohapatra:1978fy, *Georgi:1978xz,
*Barr:1979as, *Nelson:1983zb, *Barr:1984qx, *Kim:1986ax, *Peccei:1988ci,
*Cheng:1987gp}, since the only source of CP violation are the vacuum phases.
SCPV can also soften the well known SUSY CP problem~\cite{Abel:2001vy,
*Khalil:2002qp}. Finally, it is relevant to point out that in perturbative
string theory CP asymmetry can in principle only arise spontaneously through
complex VEVs of moduli and matter fields~\cite{Witten:1984dg,
*Strominger:1985it, *Dine:1992ya}.

In models of spontaneous CP violation (SCPV)
one starts from a Lagrangian that conserves CP, which implies that all
parameters of the scalar potential are real. Then, the CP
asymmetry is achieved spontaneously when the gauge interactions are broken
through complex vacuum expectation values (VEVs) of Higgs multiplets.
In fact, just having complex VEVs is not
sufficient to guarantee CP violation in the model. One has further to verify
that it is not possible to find a unitary transformation, $U$, acting on the
Higgs fields as
\begin{equation}
\phi_{i}\longrightarrow \phi_{i}^{\prime} =U_{ij}\phi_{i}\,,
\end{equation}
such that the following condition holds
\begin{equation}
\label{eq:U}
U_{ij}\, \langle \phi_{j} \rangle^{\ast} = \langle \phi_{i} \rangle \,,
\end{equation}
while leaving the full Lagrangian invariant.
If such a transformation is found, CP is a conserved
symmetry even in the presence of complex Higgs VEVs.

The main purpose of this Letter is the search for a discrete symmetry that leads
to a framework of SCPV where the VEVs of the Higgs multiplets have geometrical
values, independently of any arbitrary coupling constants in the scalar
potential - \emph{i.e.} \emph{calculable phases}~\cite{Branco:1983tn}. If such a
symmetry exists calculable phases are stable against radiative
corrections~\cite{Weinberg:1973ua,*Georgi:1974au}.
It has been shown in Ref.~\cite{Branco:1983tn} that calculable phases leading
to geometrical SCPV require more than two Higgs doublets and non-Abelian
symmetries, otherwise it is always possible to find an unitary
transformation, $U$, which is a symmetry of the potential and
fulfills Eq.~\eqref{eq:U}. The authors found an interesting example of
calculable phases with SCPV in the case of three Higgs doublets
transforming under the discrete symmetry $\mathsf{\Delta(27)}$.
In order to find symmetries that generate calculable SCPV, we start by
considering the most general
$\mathsf{SU(2)}\times \mathsf{U(1)}$ potential $V (\phi)$ with three Higgs
doublets $\phi_i$, having identical
hypercharge,
\begin{equation}
\label{eq:pot}
\begin{split}
V &(\phi) =
\sum_i \left[
-\lambda_i \phi_{i}^{\dagger} \phi_{i} + A_{i} (\phi_{i}^{\dagger} \phi_{i})^2
\right]\\&
+\sum_{i<j}
\left[
	\frac{\gamma_{i}}{2} (\phi_{i}^{\dagger} \phi_{j} + \text{H.c.})
	+ C_{i} (\phi_{i}^{\dagger} \phi_{i}) (\phi_{j}^{\dagger}
\phi_{j})\right.\\&\left.
	+\, \bar{C}_{i} \left| \phi_{i}^{\dagger} \phi_{j} \right| ^2
	+ \frac{D_{i}}{2} \left( (\phi_{i}^{\dagger} \phi_{j})^2  +
\text{H.c.}\right)
\right]\\&
+\frac{1}{2}\sum_{i \neq j}
\left[
E_{1ij}(\phi_{i}^{\dagger} \phi_{i})(\phi_{i}^{\dagger} \phi_{j}) +
\text{H.c.}
\right]\\&
+\frac{1}{2}\sum_{\substack{i \neq j \neq k\\j<k}}
\left[
E_{2i} (\phi_{i}^{\dagger} \phi_{j})(\phi_{k}^{\dagger}
\phi_{i})
+E_{3i} (\phi_{i}^{\dagger} \phi_{i})(\phi_{k}^{\dagger}
\phi_{j})\right.\\&\left.
+\,E_{4i} (\phi_{i}^{\dagger} \phi_{j})(\phi_{i}^{\dagger}
\phi_{k})\,+\,\text{\text{H.c.}}
\right]\,,
\end{split}
\end{equation}
where the constants $\lambda_i$, $A_{i}$, $\gamma_{i}$, $C_{i}$, $\bar{C}_{i}$,
$D_{i}$, $E_{2i}$, $E_{3i}$, $E_{4i}$, and $E_{1ij} \,$, $\forall_{i,j} \,,
i,j=1,2,3$ are taken real since CP invariance is imposed at the Lagrangian
level. In
what follows it is convenient to parametrize the VEVs of the doublets with
explicit phases:
\begin{equation}
\label{eq:phi_theta}
\langle\phi_1\rangle=v_1e^{i\varphi_1}\,,\quad
\langle\phi_2\rangle=v_2e^{i\varphi_2}\,,\quad
\langle\phi_3\rangle=v_3e^{i\varphi_3}\,,
\end{equation}
with the requirement of satisfying the
experimental constraint from
the heavy $W^{\pm},\,Z$ gauge boson masses:
\begin{equation}
 v^2\, \equiv\,
v_1^2+v_2^2+v_3^2\,=\,\left(\sqrt{2}\,G_F\right)^{-1},
\end{equation}
where $G_F$ is the Fermi constant.

Once the Higgs doublets are shifted according to their respective VEVs, only
some terms in the potential depend on the phases $\varphi_i$: three $\gamma_i$
terms depend on $-\varphi_{i} + \varphi_{j} \,$; six terms of the type $E_1$,
three $E_{2i}$ and three $E_{3i}$ share the same phase-dependence as the
$\gamma_{i} \,$; three $D_{i}$ terms have phase-dependence $-2 \varphi_{i} + 2
\varphi_{j} \,$; and finally, three $E_{4i}$ terms with phase-dependence
$\theta_i$,
\begin{equation}
\label{eq:theta}
\theta_i \equiv -2\varphi_{i} + \varphi_{j} + \varphi_{k} \,,
\end{equation}
where we have assumed $ i \neq j \neq k$.

Without any further assumption, there are many different coupling constants and
the only calculable phase solution to the extremum conditions is the trivial
one. We must therefore consider particular cases that can enable SCPV with
calculable phases by reducing the number of parameters either by having coupling
constants absent or related. If several coefficients are absent, then there are
non-trivial calculable phase solutions that arise from terms which must vanish
independently. For example, if only $D_i \neq 0$ there is a solution where
\begin{equation}
 -2\varphi_{i}+2\varphi_{j}= 0~\pmod{\pi}\,,
\end{equation}
for $i\neq j$. In contrast, when the coefficients are related, non-trivial
calculable phases may appear from cancellations among terms. If all quartic
coupling constants share the same value, and the $\gamma$ bi-linears vanish, the
extremum conditions admit a solution where the terms combine to make their
appropriately weighted sum vanish; the common coupling constant factors out and
calculable phases could appear - but the phases turn out to be trivial.

The most elegant way to justify the reduction of parameters of the potential in
the search for SCPV with calculable phases is by requiring invariance under
discrete non-Abelian symmetries. Therefore we consider $\mathsf{S_3}$, since it
is the smallest non-Abelian group and affects the potential given in
Eq.~\eqref{eq:pot} by forcing coefficients of the same type
to be equal. The number of
independent parameters is then reduced to ten as first proposed in
Ref.~\cite{Derman:1978rx,*Derman:1979nf}:
\begin{equation}
\label{eq:cderman}
\begin{aligned}
\lambda =&\lambda_i \,, A=A_i \,, \gamma=\gamma_i \,, C=C_i \,,
\bar{C}=\bar{C}_i \,,  D=D_i \,,\\
E_{1} &=E_{1ij}\,, E_{\alpha}=E_{\alpha i} \,, \, i,j = 1, 2,3 \,, \alpha
= 2, 3, 4 \,.
\end{aligned}
\end{equation}

Applying $\mathsf{S_3}$ invariance to the extremum conditions, solutions are
enabled with cancellations in each type of term. These solutions do not
give calculable SCPV unless there are further constraints. In the case $D=0$,
the presence of terms with $-\varphi_{i} + \varphi_{j}$ and $\theta_i$
dependence does not lead to interesting solutions even if the coupling constants
are related. In contrast, if all terms with the $-\varphi_{i} + \varphi_{j}$
dependence are absent, there is a solution
\begin{equation}
\langle \phi \rangle^{\mathsf{T}} = \frac{v}{\sqrt{3}}\, (e^{i \varphi_1}, 1, 1)
\,,
\end{equation}
with $\cos\varphi_1 = - E_4 / 6 D \,,$
which is a calculable phase only if the coupling constants $D$ and $E_{4}$ are
related in some way by the underlying theory. Obtaining such a relationship
between these coupling constants is beyond the scope of this Letter.
Finally, if only $D\neq0$ it is interesting to see that the same
solutions are obtained
that were already possible without $\mathsf{S_3}$.

In order to further reduce the parameters of the $\mathsf{S_3}$ invariant
potential, a simple addition of a cyclic $\mathsf{C_N}$ symmetry acts
by eliminating terms. In particular, $\mathsf{C_3}$ can preserve the
$E_4$-type terms while excluding every other phase-dependent term as long as
each field $\phi_i$ transforms differently under $\mathsf{C_3}$. With this
charge assignments the group $\mathsf{S_3} \ltimes (\mathsf{C_3} \times
\mathsf{C_3}) \equiv \mathsf{\Delta(54)}$ gives rise to the same potential as
the group $\mathsf{\Delta(27)}$~\cite{Branco:1983tn}, since the only
phase-dependent terms present are of the $E_4$ class - in what follows we
consider only this kind of potential. We note that the $E_4$ terms are
automatically preserved for other discrete subgroups of $\mathsf{SU(3)}$ within
the $\mathsf{\Delta(3 n^2)}$~\cite{Luhn:2007uq} and
$\mathsf{\Delta(6n^2)}$~\cite{Ishimori:2010au} families, when $n$ is a multiple
of $3$; depending on the nature of the group chosen, one has in addition to
select an appropriate representation for the Higgs multiplet and some caution is
required in order to avoid the presence of phase-dependent terms other than
$E_4$. It is natural that different groups may lead to the same Higgs
potential, however this does not apply in general to the full Lagrangian,
\emph{e.g.} when the fermions are included.

The extremum conditions can be
written in terms of the VEVs, $v_i$, and the phases
$\theta_{i}$ defined in Eq.~\eqref{eq:theta}:
\begin{subequations}
\label{eq:extremum}
\begin{align}
\frac{\partial V}{\partial v_i} =0&=
\lambda + 2 A v_i ^2 + (C+\bar{C})\, (v_j^2 + v_k^2) \\
\nonumber+\,&E_4 v_1 v_2 v_3\cos\theta_i\,,\\[2mm]
\frac{\partial V}{\partial \varphi_i} =0&= -2 v_i \sin\theta_i + v_j
\sin\theta_j + v_k \sin\theta_k \,,
\end{align}
\end{subequations}
with the restriction $i\neq j\neq k$. From Eq.~\eqref{eq:extremum} we derive
that the only VEVs with
calculable phases are those presented in
\cite{Branco:1983tn}:
\begin{subequations}
\begin{align}
\label{eq:sola}
\langle \phi \rangle^{\mathsf{T}} &= \frac{v}{\sqrt{3}}\, (1, \omega, \omega^2)
\,,\\
\label{eq:solb}
\langle \phi \rangle^{\mathsf{T}} &= \frac{v}{\sqrt{3}}\, (\omega^2,1,1) \,,
\end{align}
\end{subequations}
with the phase $\omega \equiv e^{2 \pi i / 3}$ and up to cyclical
permutations. The
solution given in Eq.~\eqref{eq:solb} is a better candidate for SCPV since it is
not removable by any symmetry of the potential, whereas the solution given in
Eq.~\eqref{eq:sola} is.

%\section{Fermion masses and mixings}

The structure of the Yukawa couplings is determined by the fermion assignments
and is restricted by the allowed contractions. In this work we restrict
ourselves to renormalizable operators: $Q \widetilde{\phi} u^c$ and $Q \phi d^c$
($\widetilde{\phi}\equiv i \sigma_2 \phi^\ast$). Allowing higher order operators
inevitably influences the scalar potential with the square of the $E_4$ term,
even though it is present at a much higher order with four additional field
insertions. We investigate what happens when placing the fields $\phi_{i}$ in an
irreducible $\mathsf{\Delta(27)}$ or $\mathsf{\Delta(54)}$ triplet which
acquires the VEV given in Eq.~\eqref{eq:solb} and without loss of generality we
place $\phi$ in the $\mathbf{3^{1}}$ or $\mathbf{3_1^1}$ representation,
respectively.

In $\mathsf{\Delta(27)}$ we can assign the quark doublets $Q_i$ as a
$\mathbf{3^1}$ triplet or as the conjugate triplet representation
$\mathbf{3^2}$, or as a combination of the nine possible
singlets~\cite{Luhn:2007uq}. We consider simultaneously the first two choices:
one of the mass structures will always be given by the $\mathbf{3^a} \times
\mathbf{3^a} \times
\mathbf{3^a}$ invariant ($a=1,2$), with the respective right-handed (RH) quarks
forced to
be the same triplet $\mathbf{3^a}$ as $Q_i$, and is unable to accommodate the
observed
fermion hierarchy between the two heaviest generations - this choice had already
been pointed out as non-viable in \cite{Branco:1983tn}. When the $Q_i$ are
singlets, the RH quarks are then forced to be triplets. The structure for
both sectors is then $\mathbf{3^1} \times \mathbf{3^2} \times \mathbf{1_{r,s}}$
($r,s=0,1,2$). The distinct rows
of the mass matrix arise from their respective singlets, and both quark sectors
share the same type of structure. Within this class we can assign all three
$Q_i$ to a single representation (choice I), only one of the generations have a
different singlet (choice II), or all three generations are different singlets
(choice III) - this leads respectively to rank 1 mass matrices with only one
non-zero eigenvalue, the decoupling of one generation, or diagonal matrices with
three distinct eigenvalues. Notice that $M_u$ and $M_d$ must share the same
structure (e.g. both I or II), since the choice of $\mathsf{\Delta(27)}$
singlets is made on the $SU(2)$ doublets $Q$. The rank~1 solution of structure I
accounts for the hierarchy of the third generation, and the decoupling case of
structure II is also promising as a leading order structure. Both structures
provide good first order approximations to the observed fermion hierarchy.

We turn now to $\mathsf{\Delta(54)}$, where we assign $Q_i$ to its
representations: two pairs of conjugate
triplets $\mathbf{3_1^{1,2}}$, $\mathbf{3_2^{1,2}}$; four doublets
$\mathbf{2_{1,\dots,4}}$ and the
singlets $\mathbf{1}$, $\mathbf{1'}$~\cite{Ishimori:2010au}. If $Q_i$ is chosen
to be a
triplet, one quark sector has the mass matrix that arises from
three-triplet invariants such as $\mathbf{3_{1}^{a}} \times
\mathbf{3_{1}^{a}} \times \mathbf{3_{1}^{a}}$, $\mathbf{3_{1}^{a}} \times
\mathbf{3_{2}^{a}} \times \mathbf{3_{2}^{a}}$ or $\mathbf{3_{1}^{a}} \times
\mathbf{3_{1}^{a}}
\times \mathbf{3_{2}^{a}}$. All these products lead to a structure with two
degenerate eigenvalues, but the third product has the non-degenerate eigenvalue
vanish. Even when $Q_i$ is a combination of doublet and singlet, the structures
again have degenerate states. If the singlet is mismatched with the singlet of
the two-triplet product, the non-degenerate value vanishes.
Finally when $Q_i$ is a combination of singlets, the structures are rank 1
regardless of choice of singlets and these are the only promising
assignments.
We note that in an extension to the leptonic
sectors, $\mathsf{\Delta(54)}$ naturally enables charged leptons with an
hierarchical structure and neutrinos with two generations degenerate.

To summarize, in the three Higgs doublet scenario $\mathsf{\Delta(27)}$ and
$\mathsf{\Delta(54)}$ are the smallest groups that lead to complex VEVs with
calculable phases stable against radiative corrections~\cite{Branco:1983tn}.
Geometrical SCPV requires the three Higgs to be assigned as a triplet
of the respective groups. Within this framework, we have explicitly investigated
the
possible fermion mass matrices of all classes of renormalizable models. In some
cases it is possible to simultaneously obtain promising first order
approximations to the observed patterns of fermion masses and mixings of
the up and down quark sectors.
In conclusion, spontaneous CP violation with calculable phases may be viable,
and a necessary condition is the correct interplay between the scalar content
and an appropriate non-Abelian symmetry group.
Since the number of physical scalar states is increased, one expects a richer
phenomenology that could be accessible
at high energy experiments, such as the Large Hadron Collider.

\begin{acknowledgments}
We thank Gustavo C. Branco for helpful discussion, reading the
manuscript and Jean-Marc G\'erard for encouragement.
This work was partially supported by Funda\c c\~ao para a Ci\^encia e a
Tecnologia (FCT, Portugal) through the projects CERN/FP/83503/2008 and CFTP-FCT
UNIT 777 which are partially funded through POCTI (FEDER) and by the Marie Curie
RTN MRTN-CT-2006-035505. The work of IdMV was supported by FCT under the
grant SFRH/BPD/35919/2007 and by DFG grant PA 803/6-1.
\end{acknowledgments}

\bibliography{refs}

\end{document}